# Time Series Analysis of Electricity Price and Demand to Find Cyber-attacks using Stationary Analysis

1st Mohsen Rakhshandehroo
*Electrical and Computer Engineering*
*Shiraz University*
Shiraz, Iran
mohsenrakhshandehroo@gmail.com

2nd Mohammad Rajabdorri
*Electrical and Computer Engineering*
*Shiraz University of Technologies*
Shiraz, Iran
rajabdorri@gmail.com

*Abstract*—With developing of computation tools in the last years, data analysis methods to find insightful information are becoming more common among industries and researchers. This paper is the first part of the times series analysis of New England electricity price and demand to find anomaly in the data. In this paper time-series stationary criteria to prepare data for further times-series related analysis is investigated. Three main analysis are conducted in this paper, including moving average, moving standard deviation and augmented Dickey-Fuller test. The data used in this paper is New England big data from 9 different operational zones. For each zone, 4 different variables including day-ahead (DA) electricity demand, price and real-time (RT) electricity demand price are considered.

*Index Terms*—

I. INTRODUCTION

Electricity grids are required to operate under reliability and integrity of data. With constant increasing electricity usage and need for smart grids development, ensuring data safety is another biggest challenge of system operation (SOs). Ahmadian [1] demonstrated a new model to participate in the market considering a nodal price limitation, which respects to the security of the data driven from the grid. According to Ray [2] risk management of the data is another critical issue with the cyber-security objective, which results in the resiliency impact of the system. Another example of the secured data driven from renewable energies is the assurance of the weather impacted data, which might result in noise and un-accuracy of the model [3]. Cyber-attacks toward electricity grids are reported in many researches [4]-[6] and therefore, different methods are presented to detect or prevent cyber-attacks, toward energy infrastructure. In [7] - [11], the attacks to automatic generation controls are investigated and counter-measurement actions using generation control diagrams are presented. Implementing statistical and data analysis methods, different detection algorithms are presented in [12]. in [13] using anomaly detection methods cyber-attacks are explored. In [14] support vector machines are used to find false data injection (FDI) attacks toward smart grids. In [15] cyber-attacks using generative adversarial networks (GANs) are presented and counter measurement actions are introduced via zero-sum game theory [16] between cuber attackers and So. The attacks toward electricity network might be due to monetary or non-monetary goals. In the monetary goals, the attacker injects FDI to change the electricity market state variables (DA or RT electricity price) [17] - [18]. In the non-monetary attacks however, the attacker aims to disrupt the operation of power systems by malfunctioning any communication infrastructure or protecting devices in the system. In these series of papers, using the big data published on New England power system operator [19] [22] the attackers toward electricity market (monetary attacks) are investigated. Since, electricity prices and demands are function of times, thus, the times-series analysis are presented in these series of papers (part1-part 4). The first step to do any times-series analysis is to test whether time-series is stationary or non-stationary and if not, the question will rise that how can we make the data to become stationary. Statistically, a times-series is said to be stationary if and if only follows the the three following conditions:

1. Constant mean
2. Constant standard
3. Auto covariance doesn't depend on times

It must be mentioned that to do further analysis, timesseries must have the stationary conditions and therefore, in this paper and in the following parts the aforementioned criteria are investigated for the New England big data.

II. MEAN AND STANDARD DEVIATION ANALYSIS

Any times-series data consist of three main parts: 1. trend, 2. seasonality or cyclicality, 3. residual error. There are specific methods to decompose time-series into these three parts that we will talk about in the next part of our times-series analysis to find the cyber-attacks [23]. In this part, the trend for times-series is explored in terms of moving average/weighted average and moving standard deviation. In fact, moving ( or rolling in some literature) average is a powerful tool to smooth the times-series and visualize the trend in the data. Moving average (MA) or moving standard deviation (Mstd) are respectively, the mean and standard deviation of data using

"n" prior samples [24]. In another word, a window of "n" past sample is considered and average and standard deviation is calculated as follows:

$$\widehat{\mu_t}(n) = \frac{1}{n} \sum_{i=6}^{n-1} y_{t-i} \quad (1)$$

$$\widehat{\sigma_t}^2(n) = \frac{1}{n-1} \sum_{i=0}^{n-1} (y_{t-i} - \widehat{\mu_t}(n))^2 \quad (2)$$

$$\widehat{\sigma_t} = \sqrt{(\widehat{\sigma_t^2}(n))} \quad (3)$$

In some literature, exponentially weighted moving average (EWMA) is considered instead of simple moving average. The (4) shows the formula for

$$\widehat{S_t} = \begin{cases} y_1 & t=1 \\ \alpha Y_t + (1-\alpha)S_{t-1} & t>1 \end{cases} \quad (4)$$

Figs. [10]-[14] show the MA, EWMA and Mstd for electricity demand and price for the DA and RT market in the 9 zones of New England operation system. As it is shown the times-series seem to be stationary for almost all the data. However, further investigation required to make sure the stationary conditions are met [25].

One of the common approaches to make the time-series as stationary, is to transfer them into different shapes. However, the Figs. [2]-[9] show that the original data is stationary, to make the analysis more comprehensive common transformations are introduced in this part. It must be emphasized that id the data is not stationary, we would do the following transformation and then we use transformed data to do the prediction or to fit ARIMA models [26].

*A. Logarithmic Transformation*

One of the common ways of smoothing data is to transform data using logarithmic function. In the following, the logarithmic transformation and corresponding MA, EWMA and Mstd are depicted for 9 zones. [fig 10-18]

*B. Removing MA Transformation*

By removing MA, the transformed time-series would be:

$$(Transformed\ time-series) = \\ (Original\ times-series) - (MA) \quad (5)$$

Thus the corresponding results would be gained:[fig 19-27]

*C. Removing EWMA transformation*

Equation. (6) shows this transformation:

$$(Transformed\ time-series) = \\ (Original\ times-series) - (EWMA) \quad (6)$$

Figures. [28]-[36] show the corresponding new time-series implementing (6).

*D. Removing Logarithmic MA Transformation*

This transformation is like (5) but both original data and MAs are in the logarithmic transformation as below:

$$log(new\ time-series) = \\ log(Original\ times-series) - log(MA) \quad (7)$$

III. AUGMENTED DICKY-FULLER TEST

In this part, the simple Dicky-Fuller and also Augmented Dicky-Fuller (ADF) are explained. In fact, ADF one of the most powerful and common industrial tools for testing data stationarity. However, MA, EWMA and Mstd are the easy and effective tools to test the stationarity by visualizing the data, ADF test is more scientific way of analyzing times series in terms of being stationary or non-stationary. In following, first DF and the ADF are introduced. Then, the analysis for any of 4 different transformation in the part 2 will be discussed.

*A. Dickey-Fuller Test*

Times Series (TS) generally can be divided into three categories based on their trends: 1. deterministic trend TS, 2. stochastic trend TS, 3. combination of stochastic and deterministic. In order to develop a TS model that can predict the future values, we need to make sure that TS doesn't explode as time increases. Unit root or DF test is the measure that we can apply to see TS is stationary or not. Before going further, at first we need to discuss three aforementioned categories, respectively.

*1) Deterministic Trend TS:* In this model, we assume that TS variable $y_t$ is a function of time ($y_t = f(t)$). This approximation can be linear, exponential or other types of fitting problems (piecewise linear, polynomial and etc.).

$$y_t = \alpha_0 + \beta_t + \varepsilon_t \quad (8)$$

If we calculate the $\Delta y_t$, we would have:

$$y_t - y_{t-1} = (\alpha_0 + \beta_t + \varepsilon_t) - (\alpha_0 + \beta_{t-1} + \varepsilon_{t-1}) \\ \Delta y_t = \beta + \varepsilon_t - \varepsilon_{t-1} \quad (9)$$

In (9), it is clear that:

$$E[\Delta y_t] = E[B] + E[\varepsilon_t] + E[\varepsilon_{t-1}], \\ E[\varepsilon_t] = E[\varepsilon_{t-1}] = 0 \quad (10)$$

Thus:

$$E[\Delta y_t] = B \quad (11)$$

It means that the average of corresponding time series $\Delta y_t$ is constant over the time and therefore the $\Delta y_t$ is stationary TS. Moreover, linear trend TS in (8) is also called trend stationary TS.

*2) Stochastic Trend TS:* In the type of TS modelling, it is considered that $y_t$ is a function of flagged $y_t$. The simple case is called first order. Auto regression model in which $y_t$ is only a function of first lag (k=1). This model is called ARClI:

$$y_t = \alpha_0 + \Phi y_{t-1} + \varepsilon_t \quad (12)$$

Based on DF test, it is proven that as long as $|\Phi| < 1$, TS in (12) is stationary. To investigate more, let's consider $\Phi = 1$, in this case we have:

$$y_t = \alpha_0 + \beta_{t-1} + \varepsilon_t$$

$$y_{t-1} = \alpha_0 {}^+ . \beta_{t-1} + \varepsilon_{t-1}$$

$$\quad (13)$$

$$y_0 =. y_0$$

Replacing the recursive ($y_{t-p}, p = 1,2,...,t-1$) values in $y_t$, we have:

$$y_t = t\alpha_0 + y_0 + \sum_{i=1}^{t} \varepsilon_i \quad (14)$$

As (14) shows, by increasing the $t$, the average values of $y_t$ will change and thus, the TS becomes non-stationary and unpredictable. The TS model in (12), when $\Phi = 1$ is called random walk with drift ($\alpha_0$). If $\alpha_0 = 0$, then TS is called random walk. The random walk is a process that can not be predicted, or in another word, there is no pattern in the data to predict. Mathematically speaking, random walk is:

$$y_t = y_{t-1} + \varepsilon_t \quad (15)$$

Moreover, in a random walk process (model) the best prediction for $t+1$ values is the current value plus some noise (error term). Thus, it can be concluded that the random walk is a non-stationary TS. It also a problem that the mean value of random walk is constant however, the variable is not.

*3) Unit Test or DF Test:* In a stationary system, the effect of any shock will damp, gradually. However, in a non-stationary system, the shocks have permanent influences on the system. In (12) for $|\Phi = 1|$, the problem is called to have the unit root. Meanwhile for the $|\Phi > 1|$, the TS becomes divergent or exploded when $t \to \infty$. To test whether the TS is stationary, DF test poses the Null and alternative hypothesis as follows:

$$considering \quad \Delta y_t = \alpha_0 + \gamma y_{t-1} + \varepsilon_t$$
$$\begin{cases} H_0 : \gamma = (1 - \Phi) = 0 \\ H_y = \gamma < 0 \end{cases} \quad (16)$$

To run Df test, we just need to run the simple least square regression to obtain the t-statistic for predicted value of $\gamma$ (which is considered as $\gamma$). The t-statistic for $\gamma$ is as follows: b

$$t - statistics(\hat{\gamma}) = \frac{\hat{\gamma}}{standard\ error\ of\ \hat{\gamma}} \quad (17)$$

By Using (17), the p-value can be obtained easily and then based on the specified significant level ($\alpha = 0.05$), if the p-value is less than $\alpha$, then we can reject the $H_0$ (Null hypothesis). Indeed if we assume that the $H_0$ is true, and statistical values (mean, std, t-statistic and etc.) for a sample of our population, then p-value is the probability of getting another sample (or better to say probability of generating a sample) with statistical values of what we already calculated for a given sample. If this probability is less than $\alpha$, it means that although it is assumed that probability of getting statistical values of a sample instead of values in $H_0$ is less than %$\alpha$ (%5 traditionally, but it happened for a given sample). Thus, the $H_0$ values are not trustworthy or we have enough evidence to reject the Null hypothesis. On the other hand, if the p-value is greater than $\alpha$ it means that considering $H_0$ to be true, the probability of getting another sample of data with the same values in $H_0$ is high and we can't reject the Null hypothesis. In the case of TS and Df test, if the p-value of DF test is less than significance level ($\alpha$), we reject the Null hypothesis or better to say we reject that or TS is any form of random walk and or TS is stationary.

*B. Augmented DF Test with Akaike Information Criteria (AIC)*

For the TS s with linear trends, or more complicated TSs, the first lag differences also doesn't lead to a stationary TS. Therefore, the general form of DF test using p-lags can be used to find the number of optimal lags to make a TS stationary. The general form of DF test, which also is known as ADF, is as follows:

$$\Delta y_t = \alpha_0 + \gamma y_{t-1} + \theta_1 \Delta y_{t-1} + ... + \theta_p \Delta y_{t-p} + \varepsilon_t \quad (18)$$

It must be noted that the test hypothesis is the same as before. The most important issue in the problem of finding the optimal lags is the evaluation metric for different models. It is clear that by adding more lags to the (18), $R^2$ and $R_{adj}^2$ of regression models change and therefore, they are not suitable evaluation metrics for models that have different regresses (different features). In fact, by adding more features to our regression model, we need the feature selection assessment tools to find the optimal model. ACI is the most common measurement tools in which not only considers the likelihood but also penalize the model for getting bigger (adding more features). The ACI formula is as follows:

$$ACI = -2\log(MSE) + 2(\text{degree of freedom (or)} \quad \text{number of in dependant variables}) \tag{19}$$

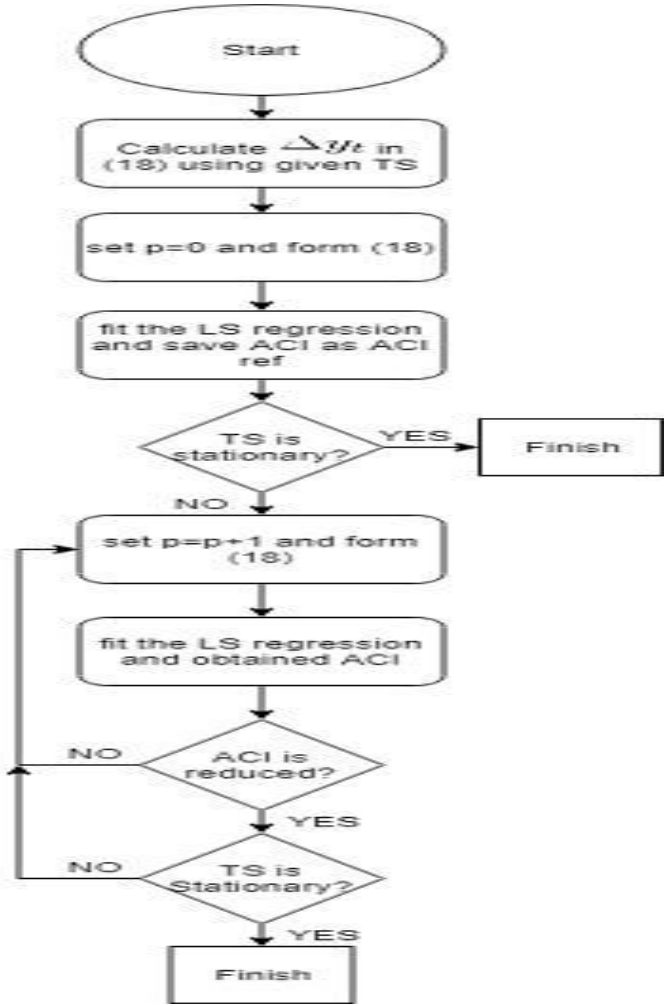

Fig. 1. ADF Test to Find the Optimal P-Lag.

The general flow chart for ADF test and to find the optimal p-lag is depicted at Fig. 10:

## IV. RESULTS FOR NEW ENGLAND BIG DATA

In this section, results of ADF test for all the TSs and their transformed version is calculated to see whether the given TS for any of 9 operational zones of NE system is non-stationary. After implementing ADF test, the next parts of this paper will address the cyber-attacks problem in electricity grids. But before doing any analysis, the first step is stationary test that we discussed in this paper.

### A. Processing All Zones' Data for DA and RT electricity price and demand

In following, different zones are analyzed based on transformations and aforementioned sections.

*1) ADF for ISONE:* Figure 2 shows the original DA and RT electricity price and demand data based on the hourly, daily, weekly and monthly time horizons for zone 1 (with code name ISONE CA).

*2) ADF for Portland:* Figure 3 shows the original DA and RT electricity price and demand data based on the hourly, daily, weekly and monthly time horizons for zone 2 (with code name Portland).

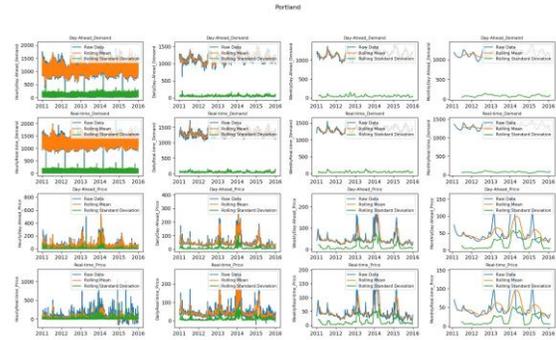

Fig. 3. Original DA and RT data for electricity price and demand at Portland.

*3) ADF for Burlington:* Figure 4 shows the original DA and RT electricity price and demand data based on the hourly, daily, weekly and monthly time horizons for zone 3 (with code name Burlington).

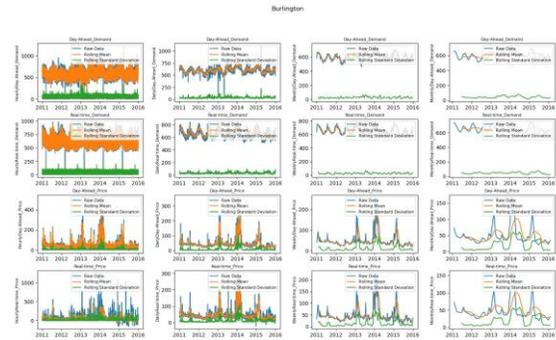

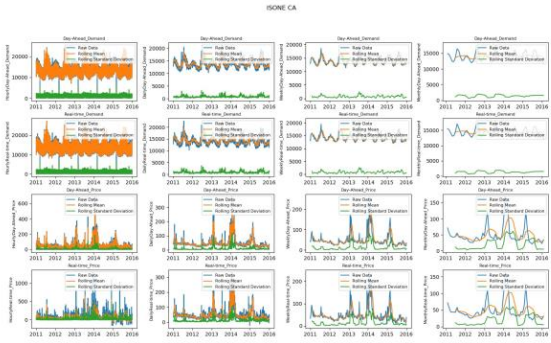
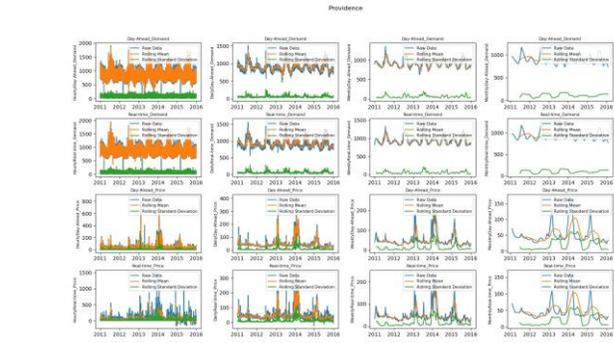

Fig. 2. Original DA and RT data for electricity price and demand at ISONE CA.
Fig. 4.	Original DA and RT data for electricity price and demand at Burlington.

Fig. 6. Original DA and RT data for electricity price and demand at Providence.

*4)	ADF for Bridgeport:* Figure 5 shows the original DA and RT electricity price and demand data based on the hourly,daily, weekly and monthly time horizons for zone 4 (with code name Bridgeport).

*8)	ADF for Boston:* Figure 9 shows the original DA and RT electricity price and demand data based on the hourly,daily, weekly and monthly time horizons for zone 8 (with code name Boston).

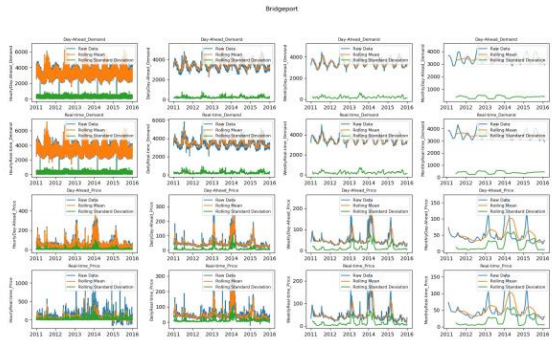
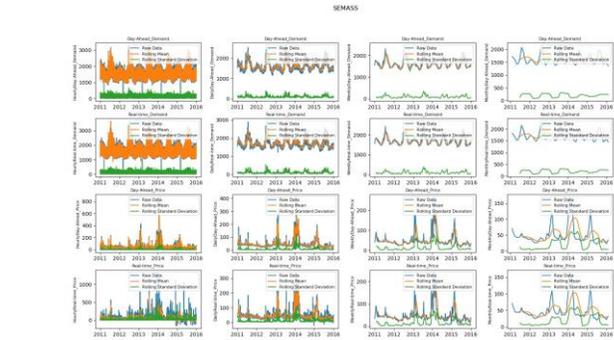

*5)	ADF for Providence:* Figure 6 shows the original DA and RT electricity price and demand data based on the hourly,daily, weekly and monthly time horizons for zone 5 (with code name Providence).

*6)	ADF for SEMASS:* Figure 7 shows the original DA and RT electricity price and demand data based on the hourly,daily, weekly and monthly time horizons for zone 6 (with code name SEMASS).

*7)	ADF for Worcester:* Figure 8 shows the original DA and RT electricity price and demand data based on the hourly,daily, weekly and monthly time horizons for zone 7 (with code name Worcester).

Fig. 5.	Original DA and RT data for electricity price and demand at Bridgeport.

To continue the analysis the RT electricity prices for first zone (ISONE CA) is investigated in following.

*B. Applying different transformation to Real-Time prices at ISONE*

In this section, the transformation filters mentioned in Section II are illustrated and compared to each other. Fig. 10 shows the linear transformation of real-time prices. After removing the MA, the data become smoother and easier to process. Fig. (11), shows the logarithmic transformation and the removed exponential moving average of the data. Fig. 12 also shows the first and second gradient of the real-time price. It is depicted clearly that these two time-series are stationary.

Fig. 13 is also the logarithmic and removed logarithmic MA of RT prices. It is shown that after removing the logarithmic MA, the TS becomes flat around the zero and it is obviously an stationary time-series.

Fig. 7. Original DA and RT data for electricity price and demand at SEMASS.

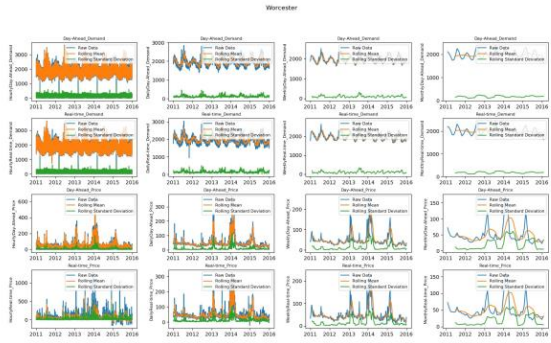

Fig. 8. Original DA and RT data for electricity price and demand at Worcester.

*C. ADF test results*

In this part the ADF results for different transformation filters are shown in TABLE I-IV. As shown, all the data from Real-Time (RT) prices and Critical-Value (CV) for 1%, 5% and 10% are categorized in ISONE CA, PORTLAND, CONCORD and BOSTON.

## V. Conclusion

In this paper, the stationary test for electricity price as well as demand is presented. The presented test uses the Augmented Dickey-Fuller (ADF) method to find the correlation between element of electricity price and demand as time-series. To identify the False Data Injection (FDI), this paper presents an efficient and fast algorithm based on ADF. Using ACI metric as the objective function of the presented algorithm, the stationary data as well as the optimal number of lags for each time-seris are obtained and can be compared with any real-time values to find mismatch and data anomaly. The results for real-time electricity price for first zone of New England system verifies the proposed method.

## Acknowledgement

All the simulations for this paper are implemented by python using the codes from Dr. S. Ahmadian github repository [20]. The authors would like to thank him for all his support and efforts for this paper.

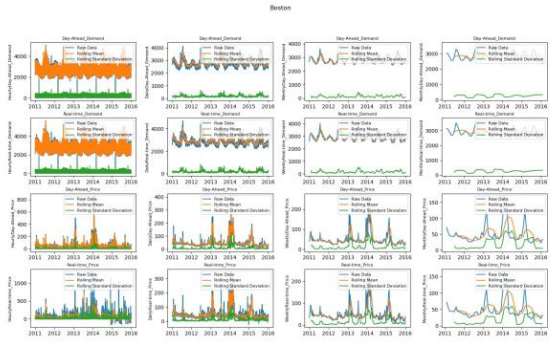

Fig. 9. Original DA and RT data for electricity price and demand at Boston.

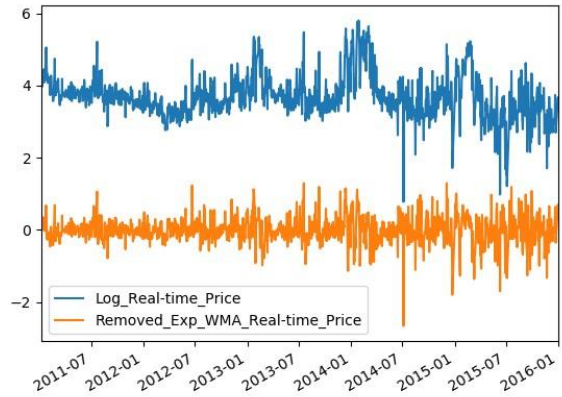

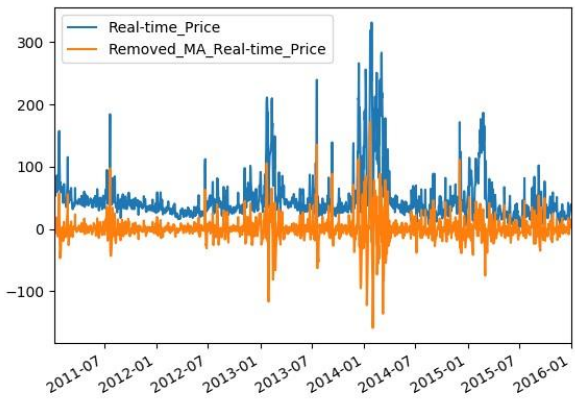

Fig. 10. Comparing real-time prices and real-time prices after removing the moving average.

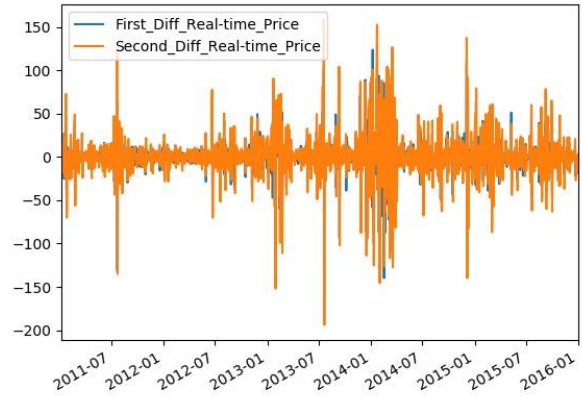

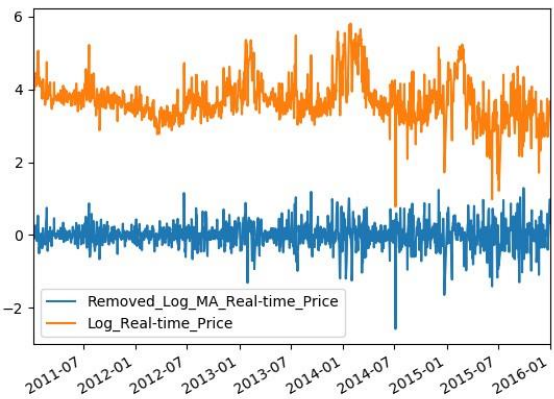

Fig. 11. Logarithmic and removed exponential moving average of the data.

Fig. 12. first and second gradient of RT prices

Fig. 13. The logarithmic and removed logarithmic MA

| | | CA DATA | | | |
| --- | --- | --- | --- | --- | --- |
| ISON | | sed | Observations | CV*(1%) | CV*(5%) | CV*(10%) |

| | Test Statistic | P-Value | #Lags Used | Observations | CV*(1%) | CV*(5%) | CV*(10%) |
|---|---|---|---|---|---|---|---|
| RT Price | -3.530843502 | 0.007229413 | 25 | 1771 | -3.4340478 | -2.863173373 | -2.567639557 |
| Log RT Price | -3.358306815 | 0.012465009 | 19 | 1777 | -3.434035296 | -2.863167853 | -2.567636618 |
| Removed MA RT Price | -8.652181977 | 5.08029E-14 | 25 | 1771 | -3.4340478 | -2.863173373 | -2.567639557 |
| Removed Exp WMA RT Price | -8.379535204 | 2.53207E-13 | 19 | 1771 | -3.434035296 | -2.863167853 | -2.567636618 |
| First Diff RT Price | -10.48431927 | 1.19125E-18 | 25 | 1771 | -3.4340478 | -2.863173373 | -2.567639557 |
| Second Diff RT Price | -9.673564151 | 1.25808E-16 | 25 | 1771 | -3.4340478 | -2.863173373 | -2.567639557 |
| Removed Log MA RT Price | -15.71760222 | 1.33595E-28 | 15 | 1781 | -3.434027007 | -2.863164194 | -2.56763467 |

TABLE II PORTLAND DATA

| PORTLAND | Test Statistic | P-Value | #Lags Used | Observations | CV*(1%) | CV*(5%) | CV*(10%) |
|---|---|---|---|---|---|---|---|
| RT Price | -3.492829162 | 0.008173178 | 25 | 1771 | -3.4340478 | -2.863173373 | -2.567639557 |
| Log RT Price | -3.399588815 | 0.010972897 | 19 | 1777 | -3.434035296 | -2.863167853 | -2.567636618 |
| Removed MA RT Price | -8.335369299 | 3.28337E-13 | 25 | 1771 | -3.4340478 | -2.863173373 | -2.567639557 |
| Removed Exp WMA RT Price | -8.497505665 | 1.26418E-13 | 19 | 1771 | -3.434035296 | -2.863167853 | -2.567636618 |
| First Diff RT Price | -10.27587346 | 3.89216E-18 | 25 | 1771 | -3.4340478 | -2.863173373 | -2.567639557 |
| Second Diff RT Price | -9.442608053 | 4.84931E-16 | 25 | 1771 | -3.4340478 | -2.863173373 | -2.567639557 |
| Removed Log MA RT Price | -14.74319096 | 2.55185E-27 | 17 | 1779 | -3.434027007 | -2.863164194 | -2.56763467 |

TABLE III CONCORD DATA

| | Test Statistic | P-Value | #Lags Used | Observations | CV*(1%) | CV*(5%) | CV*(10%) |
|---|---|---|---|---|---|---|---|
| Log RT Price | -3.388073012 | | | | | | |
| Removed MA RT Price | -8.42383204 | | | | | | |
| Removed Exp WMA RT Price | -8.460761378 | | | | | | |
| First Diff RT Price | -10.334693 | | | | | | |
| Second Diff RT Price | -9.524186778 | | | | | | |
| Removed Log MA RT Price | -14.76321501 | | | | | | |
| CONCORD | Test Statistic | P-Value | #Lags Used | Observations | CV*(1%) | CV*(5%) | CV*(10%) |
| RT Price | -3.495470938 | 0.008104175 | 25 | 1771 | -3.4340478 | -2.863173373 | -2.567639557 |
| | | 0.011372244 | 19 | 1777 | -3.434035296 | -2.863167853 | -2.567636618 |
| | | 1.95093E-13 | 25 | 1771 | -3.4340478 | | |

|  |  |  |  |  |  |  |
|---|---|---|---|---|---|---|
|  | 1.56965E-13 | 19 | 1771 | - | 2.863173373 | 2.567639557 |
|  |  |  |  | 3.434035296 | - | - |
|  | 2.78359E-18 | 25 | 1771 | -3.4340478 | 2.863167853 | 2.567636618 |
|  |  |  |  |  | - | - |
|  | 3.00852E-16 | 25 | 1771 | -3.4340478 | 2.863173373 | 2.567639557 |
|  |  |  |  |  | - | - |
|  | 2.38447E-27 | 17 | 1779 | - | 2.863173373 | 2.567639557 |
|  |  |  |  | - | - | -2.56763467 |
|  |  |  |  | 3.434027007 | 2.863164194 |  |

TABLE IV
BOSTON DATA

| BOSTON | Test Statistic | P-Value | #Lags Used | Observations | CV*(1%) | CV*(5%) | CV*(10%) |
|---|---|---|---|---|---|---|---|
| RT Price | -3.539859378 | 0.007020522 | 25 | 1771 | -3.4340478 | -2.863173373 | -2.567639557 |
| Log RT Price | -3.340968242 | 0.013143605 | 19 | 1777 | -3.434035296 | -2.863167853 | -2.567636618 |
| Removed MA RT Price | -8.687334449 | 4.12928E-14 | 25 | 1771 | -3.4340478 | -2.863173373 | -2.567639557 |
| Removed Exp WMA RT Price | -8.334585931 | 3.29853E-13 | 19 | 1771 | -3.434035296 | -2.863167853 | -2.567636618 |
| First Diff RT Price | -10.52153671 | 9.65436E-19 | 25 | 1771 | -3.4340478 | -2.863173373 | -2.567639557 |
| Second Diff RT Price | -9.725470775 | 9.29965E-17 | 25 | 1771 | -3.4340478 | -2.863173373 | -2.567639557 |
| Removed Log MA RT Price | -12.4929056 | 2.91868E-23 | 15 | 1781 | -3.434027007 | -2.863164194 | -2.56763467 |